# Fairness for ABR multipoint-to-point connections

Sonia Fahmy, Raj Jain, Rohit Goyal, and Bobby Vandalore

The Ohio State University, 2015 Neil Ave, DL 395, Columbus, OH 43210-1277, USA

## ABSTRACT

In multipoint-to-point connections, the traffic at the root (destination) is the combination of all traffic originating at the leaves. A crucial concern in the case of multiple senders is how to define fairness within a multicast group, and among groups and point-to-point connections. Fairness definition can be complicated since the multipoint connection can have the same identifier (VPI/VCI) on each link, and senders might not be distinguishable in this case. Many rate allocation algorithms implicitly assume that there is only one sender in each VC, which does not hold for multipoint-to-point cases. We give various possibilities for defining fairness for multipoint connections, and show the tradeoffs involved. In addition, we show that ATM bandwidth allocation algorithms need to be adapted to give fair allocations for multipoint-to-point connections.

**Keywords:** ATM networks, traffic management, congestion control, ABR service, multipoint communication, multipoint-to-point connections

## 1. INTRODUCTION

Multipoint communication is the exchange of information among multiple senders and multiple receivers. Multipoint support in Asynchronous Transfer Mode (ATM) networks is essential for efficient duplication, synchronization and coherency of data in such networks. Examples of multipoint applications include audio and video conferencing, server and replicated database synchronization, advertising, and data distribution applications. Multipoint-to-point connections are especially important for overlaying IP networks and simplifying end systems and edge devices.[1] In this case, only one connection needs to be set up even if there are multiple senders.

An efficient and flexible ATM multipoint service is a key factor in the success of ATM networks. Several issues need to be addressed in the ATM multipoint service definition, such as routing, reliable transport and traffic management. In this paper, we focus on traffic management issues in the case of multiple senders. Specifically, we tackle the definition of fairness, and the ABR flow control problem for multipoint-to-point connections.

ATM networks currently offer five service categories: constant bit rate (CBR), real-time variable bit rate (rt-VBR), non-real time variable bit rate (nrt-VBR), available bit rate (ABR), and unspecified bit rate (UBR). Switches generally service CBR and VBR traffic in preference to ABR traffic. The left-over capacity is *fairly* divided among the active ABR sources.[2] The most commonly adopted fairness definition is max-min fairness.[3] Intuitively, this means that all sources bottlenecked at the same node are allocated equal rates. This definition was developed for point-to-point connections, and in this paper, we attempt to extend it for multipoint connections.

For *point-to-multipoint* ABR connections, the source is usually controlled to the minimum rate supported by all the leaves of the multipoint tree, if the leaves cannot tolerate cell loss. Therefore, the extension of the max-min fairness definition to point-to-multipoint connections is straightforward. With multipoint-to-point and multipoint-to-multipoint connections, however, the implicit assumption that each connection has only one source is no longer valid.

In this paper, we define several methods for computing the max-min fair allocations for multipoint-to-point VCs, and discuss the necessary modifications to switch schemes to give these allocations. The remainder of this paper is organized as follows. The next section discusses ATM multipoint support and the solutions to the merging and cell interleaving problem for multipoint connections. Then, previous work on multipoint-to-point algorithms is summarized. We present our max-min fairness definitions in section 5, and show their operation, merits and drawbacks with the aid of examples. We then discuss several design issues (section 6), and examine how switch schemes need to be adapted to give max-min fair allocations in section 7. The paper concludes with a summary of the issues and tradeoffs involved.

Further author information:
Tel: (614) 688-4482; Fax: (614) 292-2911; E-mail: {fahmy, jain}@cis.ohio-state.edu; WWW: http://www.cis.ohio-state.edu/~jain/

## 2. ATM MULTIPOINT SUPPORT AND CELL INTERLEAVING SOLUTIONS

ATM multipoint communication is currently being studied at the ATM Forum and at the International Telecommunications Union (ITU). The Internet Engineering Task Force (IETF) has also studied the mapping of IP multicast to ATM networks. ATM user to network interface (UNI) signaling currently supports multicast via *point-to-multipoint VCs*. The ATM UNI 3.1 signaling standard supports the source-based tree approach for multicast data distribution, and uses root-initiated joins for multicast tree construction. Thus, only the root can setup the point-to-multipoint connection, add leaves, and send ATM cells. Receiver or leaf initiated join (LIJ) is a more scalable approach since it avoids this bottleneck, so UNI 4.0 signaling supports such joins. Pure multipoint-to-point and multipoint-to-multipoint services are not yet supported, although the PNNI working group at the ATM Forum is working on the definition of multipoint-to-point connections for the UBR service category. A number of proposals for supporting ATM multipoint connections as a single shared tree can be found in [4-6].

In ATM networks, the virtual path identifier (VPI)/virtual connection identifier (VCI) fields in the cell header are used to switch ATM cells. The ATM adaptation layer (AAL) at the sender segments packets into ATM cells, marking the last cell of each packet. The AAL at the receiver uses the VPI/VCI fields and the end of packet marker to reassemble the data from the cells received.

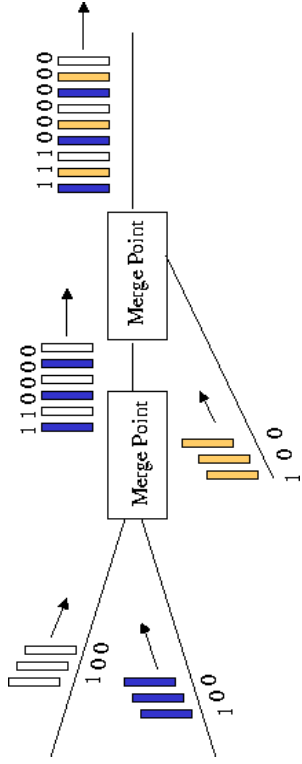

**Figure 1.** The cell interleaving problem

ATM adaptation layer 5 (AAL5), which is used for most data traffic, does not introduce any multiplexing identifier or sequence number in ATM cells. If cells from different senders are merged and interleaved on the links of a multipoint connection (implemented as a shared tree), the AAL5 at the receiver cannot assemble the data. This is because all traffic within the group uses the same VPI/VCI. The AAL5 layer uses the end-of-message bit to determine the end of each packet, but since the cells of different packets are interleaved, all the packets may get corrupted, as illustrated in figure 1. The identity of the sender is not indicated in each cell. Hence, alternate solutions must be implemented.

Most solutions to this problem attempt to either entirely avoid merging, or to prevent interleaving of cells of packets originating from different sources on the same multipoint connection after merging, or to provide enough information in the cell headers to enable the receivers to reassemble the packets even if their cells are interleaved.

The solutions proposed to the cell interleaving problem include:

1. **AAL3/4:** AAL3/4 can be used instead of AAL5. AAL3/4 contains a 10-bit multiplexing identifier (MID) field, part of which can be used to distinguish the senders in the multipoint VC. This can make switching fast and connection management simple. However, AAL3/4 suffers from excessive overhead and is not well supported. An alternative AAL, AAL5+, was proposed in [4].

2. **VC mesh:** Another solution is to overlay one-to-many VCs to create many-to-many multicast, forming a VC mesh.[7] In this case, cells from different senders can be differentiated based on their VPI/VCI fields. This solution does not scale and requires $N$ one-to-many VCs for $N$ senders.

3. **Multicast servers (MCSs):** In this case, all senders send to the MCS, which forwards data on a point-to-multipoint VC.[8] This approach is simple. The problem with it is that it is inefficient, and the MCS needs large amounts of buffering. In addition, the MCS can become a single point of congestion, which makes it difficult to guarantee quality of service requirements.

4. **Election:** Election can be used to coordinate senders. In this approach, a sender must acquire a control message before it can transmit data, and there is only one token in each VC. Hence, only one sender can transmit at a time, and no cell interleaving can possibly occur. This approach is used in the SMART scheme.[5] Although this mechanism is feasible, the overhead and delay of the scheme are high.

5. **VC merge:** The VC merge approach buffers cells of other packets at the switch until all cells of the current packet go through (as shown in figure 2). The technique is also called "cut-through forwarding," and it is used in the SEAM[6] and ARIS schemes. It entails the implementation of a packet-based scheduling algorithm at the merging point, and maintaining separate queues for each sender. The AAL5 end-of-message bit is used to signal to the switch that a packet from a different port can now be forwarded. The approach is extremely fast and simple, but it may require more memory at the switches, and add to the burstiness and latency of traffic. An analysis in [9] shows that both of these effects are minimal.

6. **VP merge:** This approach uses multipoint virtual paths (VPs). Only the VPI field is used to uniquely identify the sender. Connection management is simple in this case, but the approach requires receivers to have static assignment of VCs within VPs. In addition, VPs should not be used by end-systems, since network providers use VPs for aggregation in the backbone. Finally, there are only $2^{12} = 4096$ unique VPI values possible at each hop, and hence it is possible to run out of VPI values.

7. **Variable VP merge:** Different VPI field sizes are used in this approach.[10] The switches support both 12-bit VPI fields, as well as 18-bit VPI fields. Distributed schemes to assign globally unique VCIs within each VP are proposed using collision avoidance. This approach overcomes the VP scarcity problem of VP merge, but still has the problem of using VPs. Furthermore, it complicates the switch design since two distinct VP tables need to be maintained.

8. **Sub-channel multiplexing:** A sub-channel is a "channel within a VC." Each sub-channel can be assigned an identifier called the sub-channel number to distinguish between multiple sub-channels in a VC.[11] Four bits from the Generic Flow Control (GFC) bits in the ATM cell header can carry this number. Each burst of cells is preceded by a "start" resource management (RM) cell, and followed by an "end" RM cell. The sub-channel is allocated on the "start" cell and released on the "end" cell. Sub-channel identifiers can change at every switch. This approach allows dynamic sharing by using on-the-fly mapping of packets to sub-channels. However, four bits only allow up to fifteen concurrent senders (sub-channel number hexadecimal FF indicates an idle sub-channel). If no sub-channel is available, the burst of cells is lost, so this solution may not be scalable.

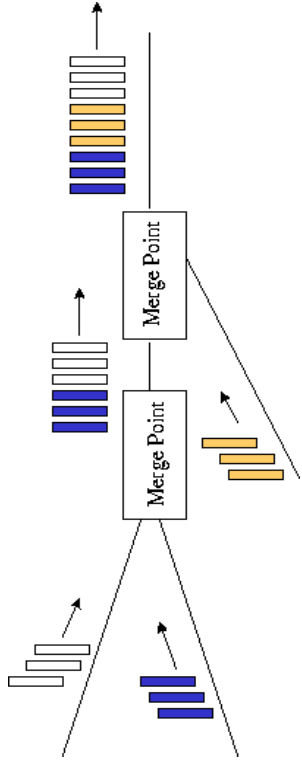

**Figure 2.** The VC merge approach

## 3. ABR FLOW CONTROL

The VC merge and VP merge approaches are the most popular approaches. This paper emphasizes the issues involved if VC merge and VP merge are implemented.

The ABR service frequently indicates to the sources the rate at which they should be transmitting. The feedback from the switches to the sources is indicated in Resource Management (RM) cells which are generated periodically by the sources and turned around by the destinations. Figure 3 illustrates this operation.

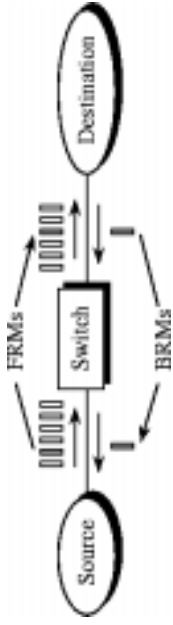

**Figure 3.** Resource management cells in an ATM network

The RM cells contain the source current cell rate (CCR), in addition to several fields that can be used by the switches to provide feedback to the sources. Among these fields, the explicit rate (ER) field indicates the rate that the network can support for this connection at that particular instant. At the source, the ER field is initialized to a rate no greater than the PCR (peak cell rate). Each switch on the path from the source to the destination reduces the ER field to the maximum rate it can support.[12]

A component $c_j$ is said to be *downstream* of another component $c_i$ in a certain connection if $c_j$ is on the path from $c_i$ to the destination. In this case, $c_i$ is said to be *upstream* of $c_j$. The RM cells flowing from the source to the destination are called forward RM cells (FRMs) while those returning from the destination to the source are called backward RM cells (BRMs). When a source receives a BRM cell, it computes its allowed cell rate (ACR) using its current ACR value, and the ER field of the RM cell.[13]

### 3.1. Fairness

The optimal operation of a distributed shared resource is usually given by a criterion called the *max-min allocation*.[3] This fairness definition is the most commonly accepted one, though other definitions are also possible.

The max-min allocation is defined as follows. Given a configuration with $n$ contending sources, suppose the $i^{th}$ source is allocated a bandwidth $x_i$. The allocation vector $\{x_1, x_2, \ldots, x_n\}$ is feasible if all link load levels are less than or equal to 100%. Given an allocation vector, the source that is getting the least allocation is, in some sense, the "unhappiest source". We need to find the feasible vectors that give the maximum allocation to this unhappiest source. Now we remove this "unhappiest source" and reduce the problem to that of the remaining $n-1$ sources operating on a network with reduced link capacities. Again, we find the unhappiest source among these $n-1$ sources, give that source the maximum allocation and reduce the problem by one source. We repeat this process until all sources have been allocated the maximum that they can get.

## 4. RELATED WORK

Little work has been done to define traffic management rules for multipoint-to-point connections. Multipoint-to-point connections require feedback to be returned to the appropriate sources at the appropriate times. As illustrated in figure 4, the bandwidth requirements for a VC after a merge point is the sum of the bandwidths used by all senders whose traffic is merged. This is because the aggregate data rate after a merging point is the sum of all incoming data rates to the merging point.[14] Similarly, the number of RM cells after merging is the sum of those from different branches. Hence, the ratio of RM to data cells remains the same.

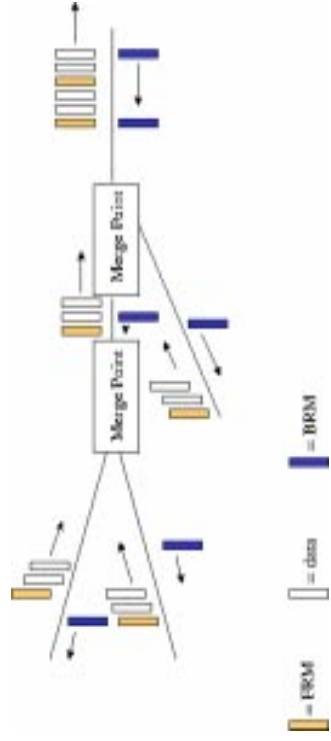

**Figure 4.** Multipoint-to-point connections

Ren and Siu[15] describe an algorithm for multipoint-to-point congestion control, which allows senders belonging to the same connection to send at different data rates. The algorithm assumes that a multipoint-to-point VC is defined as a shared tree, and that VC merging is employed to prevent the cell interleaving problem. The authors proved that if the original point-to-point switch algorithm is max-min fair, the multipoint-to-point version is also max-min fair *among sources* (and not VCs).

The idea of Ren and Siu's algorithm is very similar to point-to-multipoint algorithms (see [16,17]). The algorithm operates as follows. When a forward resource management (FRM) cell originating at a leaf is received at the merging point, it is forwarded to the root, and the merging point returns a backward resource management (BRM) cell to the source which had sent the FRM cell. The explicit rate in the BRM cell is set to the value of a register called MER (explicit rate), maintained at the merging point for each VC. The MER register is then reset to the peak cell rate. When a BRM cell is received at the merging point, the ER value in the BRM is used to set the MER register, and the BRM cell is discarded.

Another alternative is to maintain a bit at the merge point for each of the flows being merged.[18] The bit indicates that an FRM has been received from this flow after a BRM had been sent to it. Therefore, when an FRM is received at the merging point, it is forwarded to the root and the bit is set, but the RM cell is not turned around as in the previous algorithm. When a BRM is received at the merging point, it is duplicated and sent to the branches that have their bit set, and then the bits are reset. This saves the overhead that the merge point incurs when it turns around RM cells, since only destinations turn around RM cells in this case.[18]

## 5. FAIRNESS FOR MULTIPOINT-TO-POINT CONNECTIONS

In this section, we define different types of fairness, and show examples of their operation. In addition, we discuss the merits and drawbacks of each type.

### 5.1. Fairness Definitions

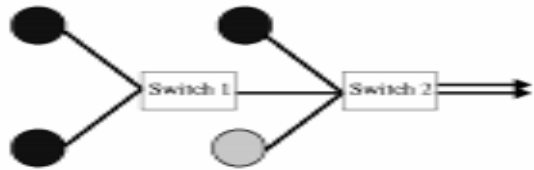

**Figure 5.** Source versus VC versus flow

Before giving the fairness definitions, we first distinguish among sources, VCs and flows. Figure 5 shows a configuration with 2 VCs. One of the VCs is a point-to-point VC, while the other is a multipoint-to-point VC. The senders in the multipoint-to-point VC are indicated by dark-colored circles, while the sender in the point-to-point VC is denoted by the light-colored circle. At the second switch, traffic from 4 sources, but only 2 VCs, is being switched to the output port. Note, however, that the second switch can distinguish 3 input flows (the point-to-point sender and 2 flows of the multipoint-to-point connection). The 2 sources whose traffic was merged at the first switch constitute a single flow at the second switch, since they cannot be distinguished downstream of their merge point. Two of the input flows that can be distinguished at the second switch belong to the same VC, while the third flow belongs to a different VC. The second switch merges the two flows of the same VC.

If a single $N$-to-one connection is treated as $N$ one-to-one connections (VCs), the max-min fairness definition can be easily extended to achieve fairness among **sources**, regardless of which VC each source belongs to. We call this source-based fairness. Note that if multipoint VCs employ the same VPI/VCI for each multipoint conversation on a certain hop, and implement VC merge at the switches, there is no way for a switch to determine the number of sources in the same multipoint VC, or maintain any type of per-source accounting information.

Observe, however, that with source-based fairness, VCs that have a larger number of concurrently active senders get more bandwidth than VCs with less concurrent senders on the same link. Thus the resource allocation is not max-min fair **among the VCs**. If VC-based max-min fairness is required, then bandwidth allocation must be max-min among VCs, and allocations to the sources in the same VC can be max-min fair within the VC. This can be done in several ways that will be explained.

A third possibility is **flow-based** max-min fairness. Intuitively, each VC coming on an input port (link) is considered a separate flow. Hence, two VCs coming on the same input port are considered two separate flows, and traffic coming from two different input ports on the same VC (and being merged at the switch) is also considered as two separate flows. *The key point is that a switch can easily distinguish the flows.*

Formally, we define a *flow* for an output port as the sum of the number of VCs sending to this output port, for each of the input ports of the switch:

$NumFlows_j, j \in OutputPorts =$
$\forall i, i \in InputPorts, \sum_i$ Number of VCs coming on port $i$ and being switched to port $j$

For example, if traffic is coming from three different input ports (ports 1, 2, and 3) and is being switched to the same output port (port 4), and one of the input ports (port 2) has two VCs sending to port 4, while each of the other two ports (ports 1 and 3) has only one VC sending to port 4 (may be the same VC, but different senders), then the number of flows at port 4 would be considered as 2 (port 2), plus 1 (port 1), plus 1 (port 3), equals four. The flow-based max-min fairness divides bandwidth fairly among the active flows. We will later see that this definition suffers from some drawbacks. The flow-based definition can be also be adopted within each VC in the VC-based approach, as seen in the next example.

The examples presented next will clarify the differences between the various fairness definitions.

## 5.2. Examples

We explain the different ways of defining fairness in multipoint situations with the aid of two examples. The first example illustrates a downstream bottleneck situation, while the second one shows an upstream bottleneck, to illustrate the allocation of capacity left-over by connections bottlenecked elsewhere.

### 5.2.1. Example 1

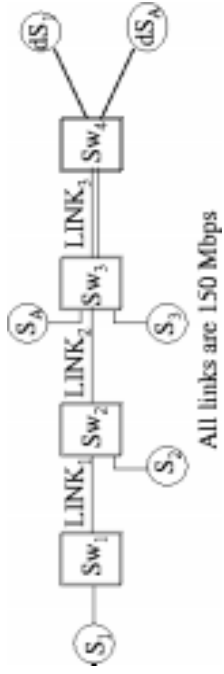

**Figure 6.** Example multipoint-to-point configuration with a downstream bottleneck

Figure 6 illustrates a configuration with two VCs: one of the VCs is a multipoint-to-point VC with three senders and one receiver, and the other is a point-to-point VC. Sources $S_1$, $S_2$, and $S_3$ are sending to destination $dS_1$, and source $S_A$ is sending to destination $dS_A$. All links are approximately 150 Mbps (after SONET overhead is accounted for). Clearly, all four sources are sharing a bottleneck link ($LINK_3$) between $Switch_3$ and $Switch_4$. **The aim of this example is to show the division of the 150 Mbps capacity of this bottleneck link among the sources.**

**Source-based Definition.** In this case, we disregard which sources belong to which connections, and simply treat this as a regular "four sources on a single bottleneck" situation. Applying the max-min fairness definition among sources, the allocations computed are:

$\{S_1, S_2, S_3, S_A\} \leftarrow \{37.5, 37.5, 37.5, 37.5\}$

Each of the four sources is allocated $\frac{1}{4} \times 150 = 37.5$.

Observe, however, that on $LINK_3$, the multipoint-to-point $VC$ is getting 3 times as much bandwidth as the point-to-point $VC$. If there were 100 concurrent senders in the multipoint-to-point $VC$, it would get 100 times as much bandwidth as the point-to-point $VC$. In essence, the bandwidth allocated to a multipoint-to-point $VC$ with $N$ concurrent senders all bottlenecked on a certain link would be $N$ times the bandwidth for a point-to-point $VC$ bottlenecked on that same link, and $N/K$ times that for a $K$-sender multipoint-to-point $VC$ bottlenecked on the same link.

**VC-based Definition: VC/Source.** If a $VC$-based definition is adopted, we are essentially dividing up the max-min fair allocation computation process into two phases. In the first phase, we ignore the number of senders in each $VC$, and simply count the $VC$s bottlenecked at each node, applying the max-min fairness computation. In the second phase, we take each multipoint-to-point $VC$ separately and divide up its allocation max-min fairly among the senders in that $VC$. This process is repeated for each multipoint-to-point $VC$.

According to this definition, the allocation vector for the example above would be:

$\{S_1, S_2, S_3, S_A\} \leftarrow \{25, 25, 25, 75\}$

This is because both of the $VC$s are bottlenecked at $LINK_3$, so each $VC$ is allocated half of the available bandwidth ($\frac{1}{2} \times 150 = 75$). Then, for the multipoint-to-point $VC$, we see that $LINK_3$ is again the bottleneck, so each of the three sources gets one third of the bandwidth allocated to this $VC$ ($\frac{1}{3} \times 75 = 25$).

**Flow-based Definition.** Recall that a flow was defined as a $VC$ coming on an input port. According to this definition, the number of flows on $LINK_3$ is three, and hence each of the flows gets one third of the bottleneck bandwidth ($\frac{1}{3} \times 150$). This bandwidth is then divided equally among the two flows seen at the output port of $Switch_2$, producing the allocation vector:

$\{S_1, S_2, S_3, S_A\} \leftarrow \{25, 25, 50, 50\}$

Clearly, this suffers from the "beat-down problem" commonly observed in EFCI situations. This means that sources whose flow travels a larger number of hops are allocated less bandwidth than those traveling a smaller number of hops, even if both flows have the same bottleneck. In flow-based fairness, sources whose flow crosses a larger number of merge points are allocated less bandwidth than those crossing a smaller number of merge points.

**VC-based Definition: VC/Flow.** If we use the $VC$-based approach, but, instead of dividing the bandwidth among the senders in the same $VC$ max-min fairly, we divide the bandwidth max-min fairly among the flows in the $VC$, a different allocation vector is obtained. For the example above, the allocation vector would be:

$\{S_1, S_2, S_3, S_A\} \leftarrow \{18.75, 18.75, 37.5, 75\}$

This is because the bandwidth is divided max-min fairly among the two $VC$s at $Switch_3$, giving 75 Mbps to each $VC$. For the multipoint-to-point $VC$, $Switch_3$ divides the 75 Mbps equally among the two flows in that $VC$, so the flow originating from $S_3$ is allocated $\frac{1}{2} \times 75 = 37.5$ Mbps. $Switch_2$ divides the 37.5 Mbps that $Switch_3$ had allocated to the flow consisting of $S_1$ and $S_2$ equally among these two sources, each getting $\frac{1}{2} \times 37.5 = 18.75$ Mbps.

### 5.2.2. Example 2

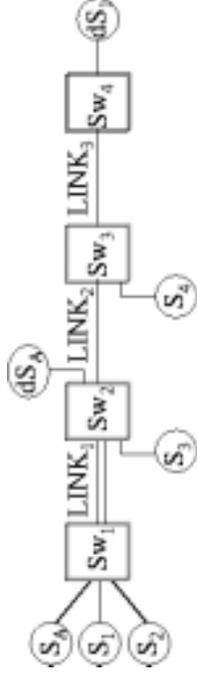

**Figure 7.** Example multipoint-to-point configuration with an upstream bottleneck

Figure 7 illustrates a configuration with two VCs: one of the VCs is a multipoint-to-point VC with four senders and one receiver, and the other is a point-to-point VC. Sources $S_1$, $S_2$, $S_3$ and $S_4$ are sending to destination $dS_1$, and source $S_A$ is sending to destination $dS_A$. All links are approximately 150 Mbps (after SONET overhead is accounted for), except for the link between $Switch_1$ and $Switch_2$ ($LINK_1$) which is only 50 Mbps. Clearly, sources $S_1$, $S_2$ and $S_A$ are bottlenecked at $LINK_1$, while sources $S_3$ and $S_4$ are bottlenecked at $LINK_3$. **The aim of this example is to illustrate the allocation of the capacity left over by sources bottlenecked on $LINK_1$ to the sources bottlenecked on $LINK_3$.**

**Source-based Definition.** The allocation vector according to the source based definition is:

$$\{S_1, S_2, S_3, S_4, S_A\} \leftarrow \{16.67, 16.67, 58.33, 58.33, 16.67\}$$

This is because each of sources $S_1$, $S_2$ and $S_A$ is allocated one third of the bandwidth of $LINK_1$. At $LINK_3$, the $50 \times \frac{2}{3} = 33.33$ Mbps used by sources $S_1$ and $S_2$ is subtracted from the available bandwidth, and the remaining capacity (116.67 Mbps) is equally divided upon sources $S_3$ and $S_4$.

**VC-based Definition: VC/Source.** According to the VC/Source definition, the allocation vector for the example above would be:

$$\{S_1, S_2, S_3, S_4, S_A\} \leftarrow \{12.5, 12.5, 62.5, 62.5, 25\}$$

This is because each of the VCs is allocated half of the bandwidth on $LINK_1$, and this bandwidth is divided equally among $S_1$ and $S_2$ of the multipoint VC. On $LINK_3$, the remaining capacity ($150 - 25 = 125$ Mbps) is divided max-min fairly among the sources within the multipoint-to-point VC.

**Flow-based Definition.** Here the allocation vector is:

$$\{S_1, S_2, S_3, S_4, S_A\} \leftarrow \{16.67, 16.67, 41.67, 75, 16.67\}$$

This is because $Switch_3$ sees two flows on $LINK_3$, and allocates half of the capacity to each flow (hence, source $S_4$ is allocated half of $LINK_3$ bandwidth). $Switch_1$ divides the 50 Mbps equally among the three flows sharing $LINK_1$ (each of $S_1$, $S_2$ and $S_A$ gets $\frac{1}{3} \times 50 = 16.67$). $Switch_2$ divides the 75 Mbps (that $Switch_3$ had allocated to the flow emerging from it) equally among the flow from $S_3$ and the flow from $Switch_1$, but detects that one of the flows (that from $Switch_1$, i.e., $S_1$ and $S_2$) is only using 33.33 Mbps, so it allocates the remaining $75 - 33.33 = 41.67$ Mbps to source $S_3$.

**VC-based Definition: VC/Flow.** According to the definition, the allocation vector for this case is:

$$\{S_1, S_2, S_3, S_4, S_A\} \leftarrow \{12.5, 12.5, 50, 75, 25\}$$

$Switch_1$ divides the available 50 Mbps equally among the two VCs (giving each 25 Mbps), and divides the bandwidth of the multipoint-to-point VC equally among the two flows in that VC (each getting 12.5 Mbps). $Switch_3$ divides the bandwidth fairly among the two flows, allocating 75 Mbps to the flow from $S_4$ and 75 Mbps to the flow from $Switch_2$. $Switch_2$ sees that the flow from $S_1$ and $S_2$ is only using 25 Mbps, so it allocates the remaining 50 Mbps to the other flow (source $S_3$).

### 5.3. Merits and Drawbacks of the Different Definitions

The two examples above illustrate how fairness based upon the concepts of source, VC, and flow give very different allocations in some situations.

First let us consider source-based fairness versus VC/source-based fairness. The source-based fairness completely ignores the membership of different sources to connections, and divides the available bandwidth max-min fairly among the sources currently active. If billing and pricing are based upon sources, it can be argued that this mechanism is good, since allocation is fair among sources.

However, if pricing is based on connections (VCs), a VC with 100 concurrent senders should not be allocated 100 times the bandwidth of a point-to-point connection bottlenecked on the same link. Source-based fairness is clearly unfair if this is the billing method adopted, and VC/source-based fairness is better.

The flow-based method is not max-min fair if we view an $N$-to-one connection as $N$ one-to-one connections, since the same flow can combine more than one source. We can, however, argue that it may be better to favor sources traversing a smaller number of merge points, since these are more likely to encounter less bottlenecks anyway. For example, if a user in New York City is fetching some web pages from a server in Germany, he expects to wait longer than if he is fetching pages from within New York City. Thus, although flow-based fairness may be unfair to sources whose traffic is merged many times with other flows, this might be acceptable in many practical situations. The VC/flow-based fairness is max-min fair with respect to VCs, but within the same VC, it favors sources whose traffic goes through a smaller number of merge points.

The above discussion shows that each type of fairness has its own merits and drawbacks, and the choice of the type of fairness to adopt relies on the billing and pricing methods used. We believe, however, that *source-based* fairness is the most preferred because it is a simple extension of point-to-point fairness definitions. To compute source-based fair allocations, a single $N$-to-one connection is treated as $N$ one-to-one connections (in terms of bandwidth allocation), regardless of which VC each source belongs to. We give (in reference [19]) a distributed algorithm that achieves source-based fairness, and show the performance analysis and simulation results of the algorithm.

The next two sections discuss the complexity of the design and implementation of algorithms to compute the above mentioned allocations.

## 6. COMMON MULTIPOINT ALGORITHM DESIGN ISSUES

There are several ways to implement multipoint-to-point ABR flow control algorithms. Each method offers a tradeoff in fairness, complexity, scalability, overhead and response time. Some of these issues are summarized next. Section 7 further discusses the implementation of multipoint algorithms.

- **VC merge versus VP merge.** With *VC merge* implementations (see section 2), it is impossible to distinguish among the cells of different sources in the same multipoint-to-point VC (since the same VPI/VCI fields are used for all the cells of a VC on the same hop). Hence, switch traffic management algorithms must not rely on being able to determine the number or rates of *active sources* with VC merge (number and rates of active VCs and number and rates of active flows can still be determined). With *VP merge*, however, the VCI field is used to distinguish among cells of different sources in the same multipoint-to-point VC on the same hop. Hence, it is possible to determine the number and rates of *active sources* in such implementations, and perform any necessary per-source accounting operations. (This, however, may incur additional complexity and reduce scalability.)

- **Per-source/VC/flow accounting.** All switch traffic management algorithms need to use some registers for storing the values they need to compute the rate allocations. Some of these values are stored for each input port, and some for each output port. Other algorithms use per-VC accounting, per-source accounting, or per-flow accounting. With multipoint-to-point VCs, per-VC accounting, per-source accounting, and per-flow accounting are no longer equivalent (they are equivalent for point-to-point scenarios). This leads to a set of interesting problems. For example, some algorithms store the value of the current cell rate (CCR) indicated in FRM cells, and later use it for computation. But the CCR value is actually per-source, and the sources cannot be distinguished with VC merge. Other algorithms also attempt to measure the source rate of senders, or distinguish between overloading and underloading sources (e.g., MIT scheme, UCSC, and ERICA schemes). This is also infeasible with VC merge. *In general, per-source accounting is infeasible with VC merge, while per-VC accounting must account for the VC as a whole (even if its traffic is coming from different ports), and per-flow accounting must distinguish both input ports and VCs.*

- **Using downstream rate allocations.** For point-to-point and point-to-multipoint connections, and for multipoint-to-point connections when using source-based fairness, the switch computes the rate allocations it can support, and then indicates these allocations in the BRM cells only if they are less than the allocations computed by downstream switches (as indicated in the ER field of BRM cells). This suffices for these situations since the algorithm operates at the source level only, and all sources at a bottleneck are allocated equal rates. With VC/source, flow, and VC/flow-based fairness, however, downstream switches compute aggregate rate allocations that must be further subdivided among senders in upstream switches. Thus, the switches must use the downstream rate allocations as an estimate of the maximum available capacity for the VC/flow.

- **BRM cell generation.** FRM cells can be turned around to BRM cells by the merge point, or by the destination. If the merge point turns around the BRM cells, the scheme may incur more overhead.

- **Scalability issues.** Some merge point algorithms wait for an FRM cell to be received before sending feedback. What are the implications of this on the scalability of the scheme? Will the feedback delay grow with the number of levels of merge points? If you have to wait for the next FRM cell at each of the merge points, the time to return a BRM cell can increase with the number of levels of the tree, which is an undesirable property. This is also dependent on the FRM cell rate, the BRM cell rate, and their relationships during transient phases. Schemes that return the BRM cell received from the root, to the leaves which have sent FRM cells to the merge point since the last BRM cell was passed, are less sensitive to number of merge points.

## 7. IMPLEMENTATION OF ALGORITHMS FOR EACH APPROACH

In section 5, we discussed four different types of fairness that can be defined for multipoint VCs. This section discusses how switch traffic management algorithms need to be adapted to compute the fair allocations for each type.

1. **Source-based fairness.**

    This type of fairness is the easiest to design and implement, since it is an extension of point-to-point algorithms. The algorithm gives the same allocation to all sources bottlenecked on the same link, and it only operates at the source level. However, source-based fairness in VC merge implementations poses some problems, since sources in the same VC cannot be distinguished. The main considerations for switch algorithms in this case is to avoid any per-source accounting and any attempt to estimate the number or rates of active sources. Note, however, that such changes may result in some oscillations and slow transient response for some algorithms, since per-source accounting and estimation of the number and rates of active sources can improve switch algorithm performance. Another consideration for schemes is to exercise special care when using the CCR field in RM cells, since the CCR value in the RM cell of a VC can belong to a different source with different bottlenecks. Reference [19] gives a switch algorithm that achieves source-based fairness and includes simulation results of that algorithm.

2. **VC/Source-based fairness.**

    VC/source-based fairness is not a straightforward extension of point-to-point algorithms, since the algorithm has to operate at two different levels: the VC level and the source level. Fair allocation of bandwidth to VCs can be simple, since VCs can be easily distinguished, their rates estimated, and the ABR available capacity can be easily measured. As with source-based fairness, VC merge implementations imply that allocations must not depend on any source-level metrics. Additional complexity is introduced by the two-level operation, which necessitates estimation of the load and capacity at both the link level and the VC level. Hence, it becomes necessary to use the explicit rates assigned by downstream switches for the VC in computing allocations at upstream switches.

3. **Flow-based fairness.**

    Flow-based fairness is non-trivial to implement. The capacity needs to be fairly divided upon the currently active flows at every node. This introduces the need for bi-level computations, since two separate flows can be merged into one flow at any node. If needed, counting the number of flows for each output link is a straightforward task, since a single bit can be maintained for each VC on each input port. If a counter maintains the number of bits set for connections to be switched to each output port, this number is the same as the number of flows on the output link. The number of active flows can be measured over successive intervals, and exponential averaging can be used to smooth out the value. Alternatively, the activity level of each flow can be estimated as the ratio of the rate of this flow and the maximum share a flow can get. The main concern for flow-based fairness, however, is that the bottleneck capacity available for a flow needs to be carefully estimated, since it depends on the explicit rate value that downstream switches allocate to the flows emerging from the switch being considered.

4. **VC/Flow-based fairness.**

      As with VC/source-based fairness, VC/flow-based fairness must operate at two different levels: the VC level and the flow-within-a-VC level. Distinguishing among VCs, and among different flows within the same VC are both quite simple. However, computing the actual allocations in a distributed manner may not be straightforward, since information from downstream switches is needed, and handling the two-level operation introduces additional complexity.

## 8. SUMMARY AND CONCLUSIONS

There are several issues to be resolved in ATM multipoint communication, including devising a scalable method for merging traffic from multiple senders, and resolving traffic management issues.

Multipoint traffic management may be implemented differently in VC merge and VP merge implementations. VP merge uses the VCI field to distinguish among different sources in the same multipoint VC, while VC merge does not distinguish sources, and implements packet-level buffering at the merge points.

Four different types of fairness can be defined for multipoint-to-point connections:

1. **Source-based fairness,** which divides bandwidth fairly among active sources as if they were sources in point-to-point connections, ignoring group memberships.

2. **VC/source-based fairness,** which first gives max-min fair bandwidth allocations at the VC level, and then fairly allocates the bandwidth of each VC among the active sources in this VC.

3. **Flow-based fairness,** which gives max-min fair allocations for each active flow, where a flow is a VC coming on an input link. Formally,
   $NumFlows_j, j \in OutputPorts =$
   $\forall i, i \in InputPorts, \sum_i$ Number of VCs coming on port $i$ and being switched to port $j$

4. **VC/flow-based fairness,** which first divides the available bandwidth fairly among the active VCs, and then divides the VC bandwidth fairly among the active flows in the VC.

Design issues common to multipoint traffic management algorithms include minimizing overhead and delays, use of VP merge versus VC merge, use of downstream allocations, and, most importantly, the use of per-source accounting, per-VC accounting and per-flow accounting in switch algorithms. Since sources, VCs, and flows are equivalent for point-to-point connections, but different for multipoint-to-point connections, it is important to note the differences between the three types of accounting. Per-source accounting cannot be performed in VC merge implementations, and can only be performed with VP merge. Per-flow accounting has to distinguish VCs and input ports, while per-VC accounting must combine the VC information coming from different input ports.

Modifications are necessary for switch algorithms to implement each of the four types of fairness. For source-based fairness (the simplest), algorithms operating with VC merge should not attempt any source-level accounting, and must only use information supplied in the RM cells, in addition to aggregate measurements of load, capacity and queuing delays. VC/source-based fairness must make VC-level allocations and source-level allocations, making use of per-VC accounting. Flow-based fairness can be achieved by estimating flow activity and available flow capacity, and VC/flow-based fairness should also estimate both VC and flow load and capacity.

It is essential to continue this work to define the desirable forms of fairness, and extend current switch traffic management algorithms for multipoint connections. Extensive performance analysis is also crucial to examine the fairness, complexity, overhead, transient response, delays, and scalability tradeoffs involved.

## ACKNOWLEDGMENTS

This research was sponsored in part by Rome Laboratory/C3BC Contract # F30602-96-C-0156.